\documentclass{moriond}

\def\Journal#1#2#3#4{{#1} {\bf #2}, #3 (#4)}

\def\NPB{{\em Nucl. Phys.} B}
\def\PLB{{\em Phys. Lett.}  B}
\def\PRL{\em Phys. Rev. Lett.}
\def\PRD{{\em Phys. Rev.} D}

\def\be{\begin{equation}}
\def\ee{\end{equation}}
\def\bea{\begin{eqnarray}}
\def\eea{\end{eqnarray}}

\newcommand{\Photo}{}

\begin{document}

%------------------------BEGINNING OF NIKHEF TITLE PAGE----------------------

\thispagestyle{empty}

\begin{flushright}
Nikhef-2018-020\\
\end{flushright}

\vspace{2.0truecm}
\begin{center}
\boldmath
\large\bf Utilising $B \to \pi K$ Decays at the High-Precision Frontier
\unboldmath
\end{center}

\vspace{0.9truecm}
\begin{center}
 Robert Fleischer\,${}^{1,2}$, Ruben Jaarsma\,${}^{1}$ \footnote{Speaker}, Eleftheria Malami\,${}^1$  and  K. Keri Vos\,${}^{3}$\\[0.1cm]

${}^1${\sl Nikhef, Science Park 105, 1098 XG Amsterdam, Netherlands}

${}^2${\sl  Department of Physics and Astronomy, Faculty of Science, Vrije Universiteit Amsterdam,\\
1081 HV Amsterdam, Netherlands}

${}^3${\sl Theoretische Physik 1, Naturwissenschaftlich-Technische Fakult\"at, \\
Universit\"at Siegen, D-57068 Siegen, Germany}
\end{center}

\vspace{2.9truecm}

\begin{center}
{\bf Abstract}
\end{center}

{\small
\vspace{0.2cm}\noindent
For about twenty years, $B\to\pi K$ decays are in the focus of $B$-decay studies. We show that a correlation between the CP asymmetries of $B^0_d\to\pi^0K_{\rm S}$ reveals a tension with the Standard Model. Should it be due to New Physics, a modified electroweak penguin sector provides particularly interesting possibilities. We present a new method to determine the electroweak penguin parameters, which uses an isospin relation and requires only minimal $SU(3)$ input. We apply it to the current data for $B\to\pi K$ decays and discuss the prospects for utilizing CP violation in $B^0_d\to\pi^0K_{\rm S}$. The strategy has the exciting potential to establish New Physics in the electroweak penguin sector in the high-precision era of $B$-physics.}

\vspace{3.9truecm}

\begin{center}
{\sl Talk given at Rencontres de Moriond 2018, QCD and High Energy Interactions\\
La Thuile, Italy, 17--24 March 2018\\
To appear in the Proceedings}
\end{center}

\vfill
\noindent
May  2018

\newpage
\thispagestyle{empty}
\vbox{}
\newpage
 
\setcounter{page}{1}

%------------------------END OF NIKHEF TITLE PAGE------------------------------

\vspace*{4cm}
\title{\boldmath UTILISING $B \to \pi K$ DECAYS AT THE HIGH-PRECISION FRONTIER}

\author{R. FLEISCHER\,${}^{1,2}$, R. JAARSMA\,${}^{1}$ \footnote[1]{Speaker}, E. MALAMI\,${}^1$ AND K. K. VOS\,${}^{3}$}

\address{${}^1$Nikhef, Science Park 105, 1098 XG Amsterdam, Netherlands \\
${}^2$Department of Physics and Astronomy, Faculty of Science, Vrije Universiteit Amsterdam,\\
1081 HV Amsterdam, Netherlands \\
${}^3$Theoretische Physik 1, Naturwissenschaftlich-Technische Fakult\"at, \\
Universit\"at Siegen, D-57068 Siegen, Germany}

\maketitle\abstracts{
For about twenty years, $B\to\pi K$ decays are in the focus of $B$-decay studies. We show that a correlation between the CP asymmetries of $B^0_d\to\pi^0K_{\rm S}$ reveals a tension with the Standard Model. Should it be due to New Physics, a modified electroweak penguin sector provides particularly interesting possibilities. We present a new method to determine the electroweak penguin parameters, which uses an isospin relation and requires only minimal $SU(3)$ input. We apply it to the current data for $B\to\pi K$ decays and discuss the prospects for utilizing CP violation in $B^0_d\to\pi^0K_{\rm S}$. The strategy has the exciting potential to establish New Physics in the electroweak penguin sector in the high-precision era of $B$-physics.}

\section{Introduction}

Decays of the type $B \to \pi K$ have been in the spotlight for over two decades (see Refs.~1,2 and references therein). This is a particularly interesting class of decays because the leading contributions come from QCD penguin topologies; the tree topologies are suppressed by the CKM matrix element $V_{ub}$. Moreover, electroweak (EW) penguin topologies give contributions at the same level as the tree topologies.

The decay $B_d^0 \to \pi^0 K_{\rm S}$ is the only $B \to \pi K$ channel with a mixing-induced CP asymmetry, which arises from interference between $B_d^0$--$\bar{B}_d^0$ mixing and the decay of $B_d^0$ or $\bar{B}_d^0$ into the $\pi^0 K_{\rm S}$ final state. Moreover, all $B \to \pi K$ decays may have direct CP violation, arising from interference between penguin and tree contributions. The correlation between the CP asymmetries of the $B_d^0 \to \pi^0 K_{\rm S}$ mode has revealed a discrepancy in the past, which could be explained by a modified EW penguin sector.\cite{FJPZ} We provide an up-to-date picture of this correlation, and present a new method to pin down the parameters governing the EW penguin contributions.\cite{FJV-1,FJMV}

\section{\boldmath The $B \to \pi K$ System}

The EW penguin topologies contributing to $B_d^0 \to \pi^- K^+$ and $B^+ \to \pi^+ K^0$ are colour-suppressed and play a minor role. On the other hand, the $B_d^0 \to \pi^0 K^0$ and $B^+ \to \pi^0 K^+$ channels have also contributions from colour-allowed EW penguin toplogies. These effects are described by the following parameter, which can be calculated using the $SU(3)$ flavour symmetry: \cite{NR,BFRS}
\begin{equation} \label{eq:qphi}
	q e^{i\phi} e^{i\omega} \equiv - \left(\frac{\hat{P}_{EW} + \hat{P}_{EW}^{\rm C}}{\hat{T} +\hat{C}}\right) \stackrel{\rm{SM}}{=} \frac{-3}{2\lambda^2 R_b}\left(\frac{C_9 + C_{10}}{C_1 + C_2}\right) R_q = (0.68 \pm 0.05) R_q.
\end{equation}
Here $\phi (\omega)$ is a CP-violating (CP-conserving) phase, and $\hat{P}_{EW} (\hat{T})$ and $\hat{P}_{EW}^{\rm C} (\hat{C})$ are colour-allowed and colour-suppressed EW penguin (tree) amplitudes, respectively. Note that $\omega$ vanishes in the $SU(3)$ limit, and that its smallness is a model-independent result.\cite{BBNS} Furthermore, $\lambda \equiv |V_{us}| = 0.22$, $R_b$ is a side of the unitarity triangle (UT), the $C_i$ are Wilson coefficients, and $SU(3)$-breaking corrections are parametrized by $R_q = 1.0 \pm 0.3$.

The tree and QCD penguin topologies are described by the hadronic parameters
\begin{equation} \label{eq:had-param}
	r_{\rm c}e^{i\delta_{\rm c}} \equiv (\hat{T}+\hat{C})/P', \quad re^{i\delta} \equiv (\hat{T}-\hat{P}_{tu})/P',
\end{equation}
where $\hat{P}_{tu}$ is the difference between QCD penguin amplitudes with $t$ and $u$ quarks, and $P' \propto P_{tc}$. In order to determine these parameters, we follow Refs.~1,2 and use $B \to \pi \pi$ data, where contributions from EW penguins are tiny, and employ the $SU(3)$ flavour symmetry, yielding: \cite{FJV-1,FJMV}
\begin{equation} \label{eq:had-param-num}
	r_{\rm c}e^{i\delta_{\rm c}} = (0.17\pm0.06)e^{i(1.9\pm23.9)^\circ}, \quad re^{i\delta} = (0.09\pm0.03)e^{i(28.6\pm21.4)^\circ}.
\end{equation}
Here, we allowed for non-factorizable $SU(3)$-breaking corrections of $20 \%$. In an analysis of $B_{d,s} \to \pi\pi, \, KK, \, \pi K$ modes we did not find indications of anomalously large non-factorizable $SU(3)$-breaking corrections. \cite{FJV-2} 

The amplitudes of the $B \to \pi K$ decays satisfy the following isospin relation: \cite{BFRS,FJPZ}
\begin{equation} \label{eq:isospin-relation-amps}
	\sqrt{2} A(B^0_d \to \pi^0 K^0) + A(B^0_d \to \pi^- K^+) = \sqrt{2} A(B^+\to \pi^0 K^+) + A(B^+\to \pi^+ K^0) \equiv 3A_{3/2}.
\end{equation}
Here $3A_{3/2} \equiv 3 |A_{3/2}|e^{i\phi_{3/2}}$ is an isospin $I=3/2$ amplitude, which is given by
\begin{equation} \label{eq:isospin-relation-qphi}
	3A_{3/2} = - (\hat{T} +\hat{C})e^{i\gamma} + (\hat{P}_{EW} + \hat{P}_{EW}^{\rm C}) = - (\hat{T} +\hat{C})\left(e^{i\gamma} - q e^{i\phi} e^{i\omega}\right),
\end{equation}
where $\gamma = (70\pm 7)^\circ$ is the corresponding UT angle, and $|\hat{T}+\hat{C}|$ can be determined from the $B \to \pi\pi$ system using the following $SU(3)$ relation: \cite{GRL}
\begin{equation} \label{eq:tplusc}
	|\hat{T}+\hat{C}| = R_{T+C}\left|V_{us}/V_{ud}\right|\sqrt{2} |A(B^+\to\pi^+ \pi^0)|.
\end{equation}
The $SU(3)$-breaking effects are given by $R_{T+C}\approx f_K/f_\pi =  1.2 \pm 0.2$, where the central value is obtained in factorization and the uncertainty allows for non-factorizable corrections.

The direct CP asymmetries $A_{\rm CP}^f \equiv \left(|\bar{A}_f|^2-|A_f|^2\right)/\left(|\bar{A}_f|^2+|A_f|^2\right)$ are proportional to $r_{({\rm c})}\sin\delta_{({\rm c})}\sin\gamma$, giving values of ${\cal O}(10 \%)$. Together with the branching ratios they are ingredients of a sum rule,\cite{G-GR} which vanishes in the SM up to corrections of ${\cal O}(r_{({\rm c})}^2)$.\cite{FJV-1,FJMV} The current experimental data \cite{PDG} are in agreement with the SM pattern.\cite{FJMV} Since the uncertainty of $A_{\rm CP}^{\pi^0 K^0}$ is still large, we use the sum rule to predict this observable: \cite{FJV-1,FJMV}
\begin{equation} \label{eq:aDir-sum-rule}
	A_{\rm CP}^{\pi^0K^0} = -0.14 \pm 0.03.
\end{equation}

The mixing-induced CP asymmetry $S^f_{\rm CP}$ enters the time-dependent rate asymmetry as
\begin{equation}
	\frac{\Gamma(\bar{B}_d^0(t) \rightarrow \pi^0K_{\rm S}) - 
\Gamma(B_d^0(t) \rightarrow \pi^0K_{\rm S})}{\Gamma(\bar{B}_d^0(t) \rightarrow \pi^0K_{\rm S}) + 
\Gamma(B_d^0(t) \rightarrow \pi^0K_{\rm S})} = A^{\pi^0K_{\rm S}}_{\rm CP} \cos(\Delta M_dt) + S^{\pi^0K_{\rm S}}_{\rm CP}\sin(\Delta M_dt),
\end{equation}
where $\Delta M_d$ is the mass difference between the $B_d$ mass eigenstates. We have
\begin{equation} \label{eq:corr-CP-asymmetries}
	S^{\pi^0K_{\rm S}}_{\rm CP} = \sin(\phi_d - \phi_{00})\sqrt{1- (A^{\pi^0K_{\rm S}}_{\rm CP})^2},
\end{equation}
where $\phi_d=(43.2\pm1.8)^\circ$ is the CP-violating $B^0_d$--$\bar B^0_d$ mixing phase.\cite{FJPZ} The key quantity is the angle $\phi_{00} \equiv \rm{arg}(\bar{A}_{00} A^*_{00})$ between $A_{00} \equiv A(B_d^0 \to \pi^0 K^0)$ and its CP-conjugate $\bar{A}_{00}$, which can be expressed in terms of the hadronic parameters in Eq.~\ref{eq:had-param} as follows: \cite{FJV-1,FJMV}
\begin{equation} \label{eq:phi00}
	\tan\phi_{00} =  2(r\cos\delta-r_{\rm c}\cos\delta_{\rm c})\sin\gamma+2r_{\rm c} \left(\cos\delta_{\rm c}-2\tilde a_{\rm C} /3 \right) q\sin\phi +{\cal O}(r_{\rm (c)}^2).
\end{equation}
Here $\tilde a_{\rm C} \equiv a_{\rm C}\cos(\Delta_{\rm C} + \delta_{\rm c})$ parametrizes the small colour-suppressed EW penguin contributions.

\section{\boldmath Correlation Between the CP Asymmetries of $B_d^0 \to \pi^0 K_{\rm S}$}

\begin{figure}
\begin{minipage}{0.5\linewidth}
\centerline{\includegraphics[width=0.58\linewidth]{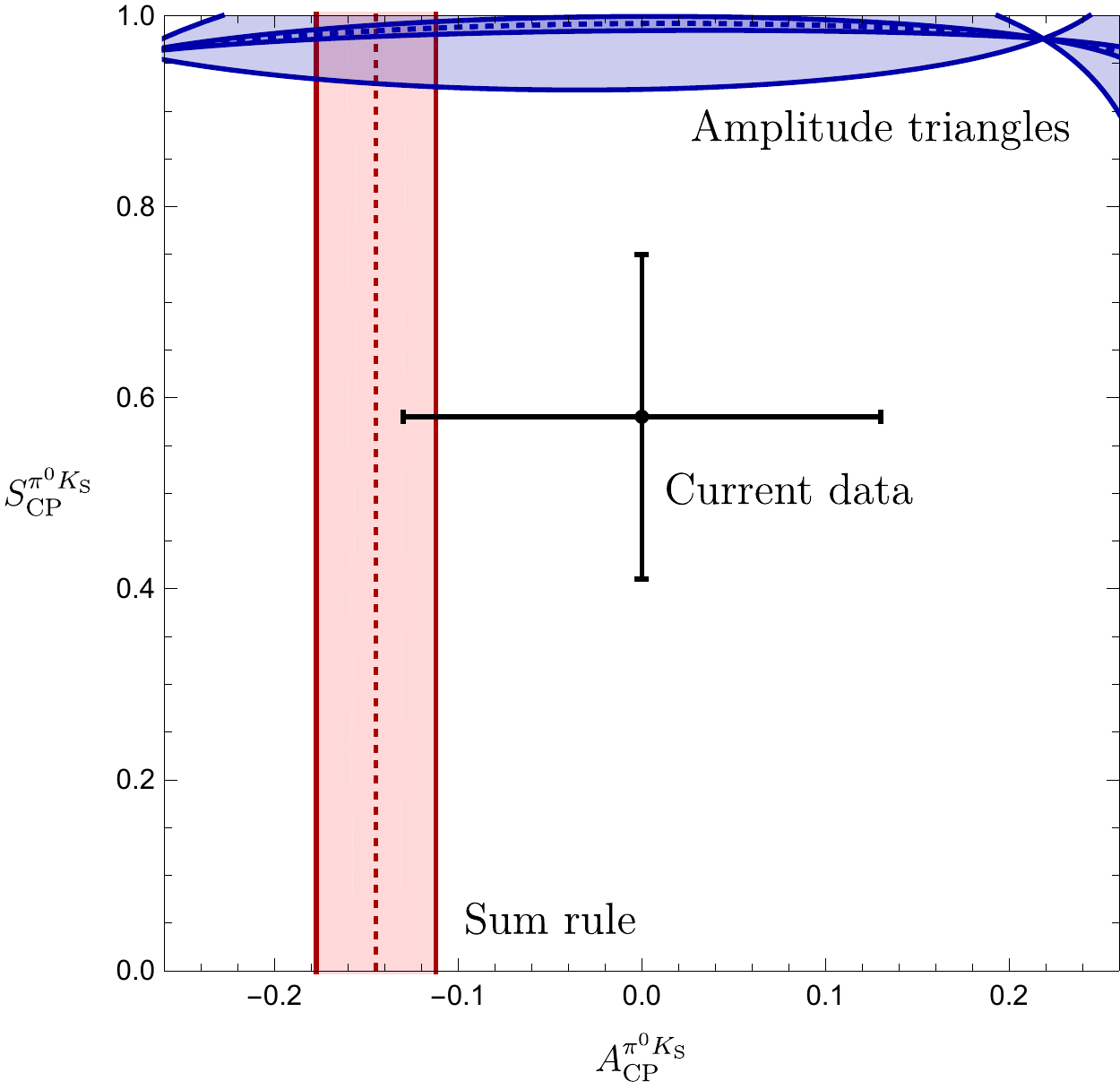}}
\end{minipage}
\hfill
\begin{minipage}{0.5\linewidth}
\centerline{\includegraphics[width=0.58\linewidth]{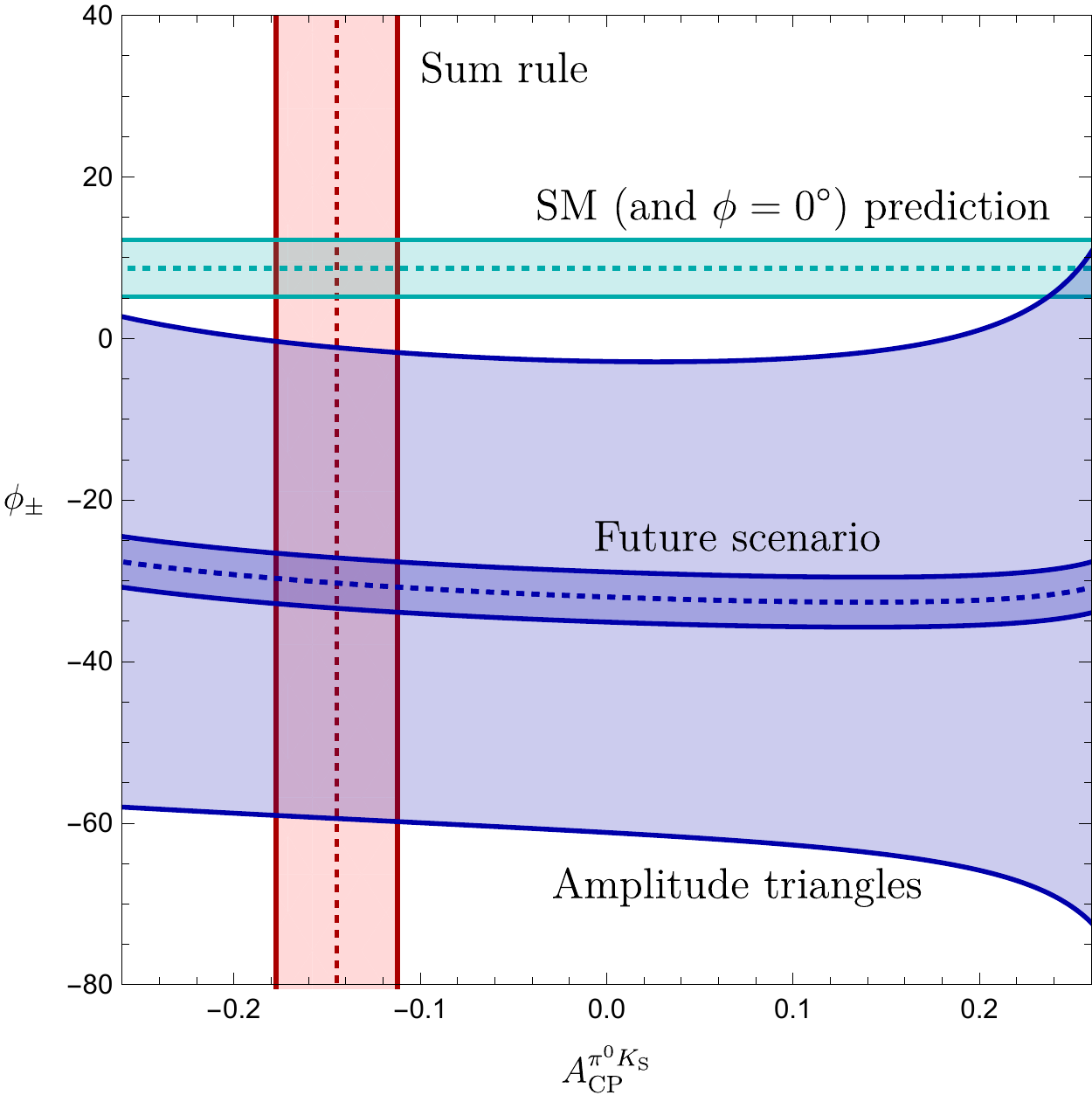}}
\end{minipage}
\caption[]{Correlation between the CP asymmetries of $B_d^0 \to \pi^0 K_{\rm S}$ (left), and $\phi_\pm$ as a function of $A_{\rm CP}^{\pi^0 K_{\rm S}}$ (right).}
\label{fig:triangle-correlation}
\end{figure}

We may calculate $\phi_{00}$ using the numerical values in Eqs.~\ref{eq:qphi}~and~\ref{eq:had-param-num}. However, the cleanest way to determine this quantity is from the amplitude triangles corresponding to the isospin relation for the neutral decays in Eq.~\ref{eq:isospin-relation-amps}, as it requires only minimal $SU(3)$ input and no topologies have to be neglected.\cite{FJPZ} From Eq.~\ref{eq:corr-CP-asymmetries}, we can then determine $S_{\rm CP}^{\pi^0 K_{\rm S}}$ as a function of $A_{\rm CP}^{\pi^0 K_{\rm S}}$. This yields the correlation shown in the left panel of Fig.~\ref{fig:triangle-correlation}, which is more constrained than in previous work~\cite{FJPZ} due to a better determination of $\gamma$. We also show current data \cite{PDG} and the prediction from the sum rule. We observe a discrepancy between the data and the correlation at the $2 \sigma$ level.

In the right panel of Fig.~\ref{fig:triangle-correlation}, we show a new constraint, obtained from the angle $\phi_\pm \equiv  \rm{arg}(\bar{A}_\pm A_\pm^*)$ between $A_\pm \equiv A(B_d^0 \to \pi^- K^+)$ and its CP-conjugate $\bar{A}_\pm$. For $\phi=0^\circ$, which includes the SM, we obtain $\left.\phi_\pm \right |_{\phi=0}= 2 \, r \cos\delta\sin\gamma + {\cal O}(r^2)= (8.7 \pm 3.5)^\circ$, where the numerical value follows from Eq.~\ref{eq:had-param-num}. We can also extract this angle from the amplitude triangles. The tension between these two constraints shows that also the correlation itself is not in agreement with the SM. We could obtain a consistent picture in Fig.~\ref{fig:triangle-correlation} if the CP asymmetries of $B_d^0 \to \pi^0 K_{\rm S}$ moved by $\sim 1\sigma$ and ${\mathcal Br}(\pi^0K^0)$ went down by $\sim 2.5 \sigma$. On the other hand, Fig.~\ref{fig:triangle-correlation} may also be a hint of NP, where a modified EW penguin sector is a particularly interesting scenario.

\section{Determination of the EW Penguin Parameters}

The EW penguin parameters $q$ and $\phi$ can also be determined from the isospin relation in Eqs.~\ref{eq:isospin-relation-amps}~and~\ref{eq:isospin-relation-qphi}. Specifically, we can use the amplitude triangles to express these parameters as a function of $3 |A_{3/2}|/|\hat{T} +\hat{C}|$, yielding contours in the $\phi$--$q$ plane.\cite{FJV-1,FJMV} This method requires only minimal $SU(3)$ input to determine $|\hat{T} +\hat{C}|$ from Eq.~\ref{eq:tplusc}, and no topologies have to be neglected.

\begin{figure}
\begin{minipage}{0.5\linewidth}
\centerline{\includegraphics[width=0.58\linewidth]{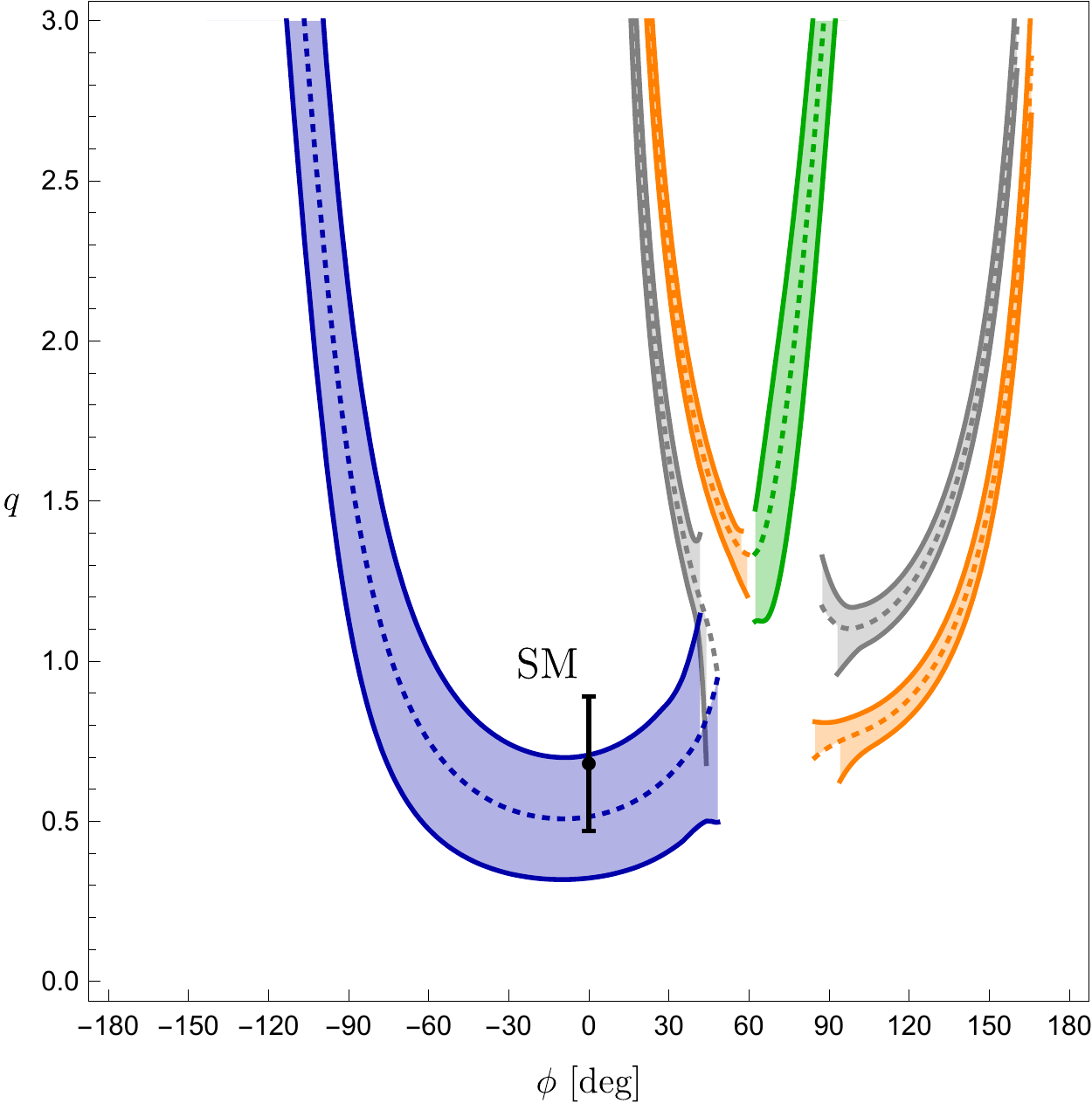}}
\end{minipage}
\hfill
\begin{minipage}{0.5\linewidth}
\centerline{\includegraphics[width=0.58\linewidth]{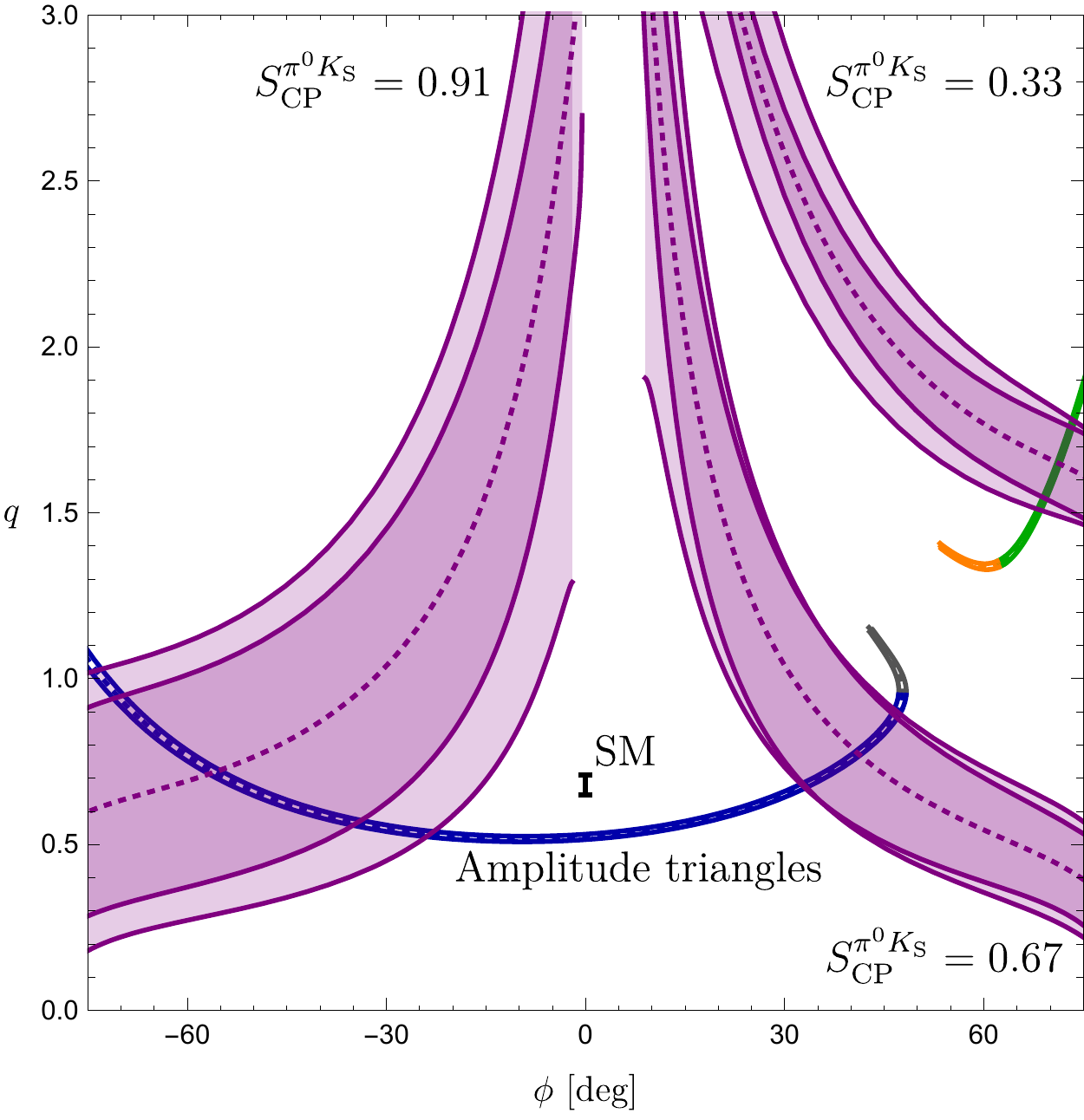}}
\end{minipage}
\caption[]{Contours in the $\phi$--$q$ plane for charged $B \to \pi K$ data following from the isospin relation in Eqs.~\ref{eq:isospin-relation-amps}~and~\ref{eq:isospin-relation-qphi}. The left panel shows current data, whereas the right one corresponds to future scenarios. The purple contours follow from different assumed measurements of $S_{\rm CP}^{\pi^0 K_{\rm S}}$.}
\label{fig:qphi}
\end{figure}

The analysis can be done for both the neutral decays and the charged decays separately. It requires us to fix the relative orientation of the triangles, which we can do through $S_{\rm CP}^{\pi^0 K_{\rm S}}$ in the case of the neutral decays, and with the angle between $A(B^+ \to \pi^+ K^0)$ and its CP conjugate, which is of ${\cal O}(1^\circ)$, for the charged decays. Since the current uncertainty of $S_{\rm CP}^{\pi^0 K_{\rm S}}$ is still large,\cite{PDG} we perform the analysis for the charged decays, yielding the contours in the left panel of Fig.~\ref{fig:qphi}.

In order to determine the values of $q$ and $\phi$ we need further input. This can be obtained from $S_{\rm CP}^{\pi^0 K_{\rm S}}$ using the hadronic parameters in Eq.~\ref{eq:had-param}. In particular, we can convert a measurement of this observables into a value of $\phi_{00}$, and subsequently obtain a contour in the $\phi$--$q$ plane from Eq.~\ref{eq:phi00}. As the strong phases enter only as $\cos\delta_{({\rm c})}$, this expression is very insensitive to variations of these small parameters, thereby having a theoretically favourable structure. Furthermore, the small contributions from colour-suppressed EW penguins can be included through data.\cite{FJV-1,FJMV}

In view of the current large uncertainty of $S_{\rm CP}^{\pi^0 K_{\rm S}}$, we study 3 different scenarios. In the right panel of Fig.~\ref{fig:qphi}, we give again the contours from the amplitudes triangles, now assuming perfect measurements and progress on the calculation of $R_{T+C}$.\cite{FJPZ} In addition, we show the contours from $S_{\rm CP}^{\pi^0 K_{\rm S}}$, where we assume a precision of $\pm 0.04$ for the CP asymmetries of $B_d^0 \to \pi^0 K_{\rm S}$ at the end of Belle II,\cite{Belle-II} and include $20 \%$ non-factorizable $SU(3)$-breaking corrections for the hadronic parameters entering Eq.~\ref{eq:phi00}. We give separately the experimental (small bands) and theoretical (wide bands) uncertainties, and observe that we can match the experimental precision with theory. Moreover, we see that $S_{\rm CP}^{\pi^0 K_{\rm S}}$ provides complementary information on $q$ and $\phi$, allowing the determination of these parameters.

\section{Conclusions}

We have performed a state-of-the-art $B \to \pi K$ analysis, showing that a tension with the SM in the correlation of the $B_d^0 \to \pi^0 K_{\rm S}$ CP asymmetries has become stronger. In order to clarify this intriguing picture, either data have to move to confirm the SM, or we may have NP, where a modified EW penguin sector provides a particularly interesting scenario. We present a new strategy to determine the EW penguin parameters $q$ and $\phi$, which has the potential to resolve this puzzling situation and reveal new sources of CP violation.

\section*{Acknowledgments}

This research has been supported by the Netherlands Organisation for Scientific Research (NWO) and by the Deutsche Forschungsgemeinschaft (DFG), research unit FOR 1873 (QFET).

\end{document}